\documentclass[numberedappendix,apj,twocolumn]{emulateapj}

\newcommand{\lya}{Ly$\alpha$}

\newcommand{\bFilter}{$B_{435}$}
\newcommand{\vFilter}{$V_{606}$}
\newcommand{\iFilter}{$i_{775}$}
\newcommand{\zFilter}{$z_{850}$}
\newcommand{\yFilter}{$Y_{105}$}
\newcommand{\ymFilter}{$Y_{098}$}

\newcommand{\jFilter}{$J_{125}$}

\newcommand{\hFilter}{$H_{160}$}

\newcommand{\Jdrop}{$J_{125}$-dropout}
\newcommand{\msol}{$M_\odot$}

\shorttitle{Expanded Search for $z\sim10$ Galaxies}
\shortauthors{Oesch et al.}

\begin{document}

\title{Expanded Search for $z\sim10$ Galaxies from HUDF09, ERS, and CANDELS Data: \\Evidence for accelerated evolution at $z>8$?
\altaffilmark{1}}

\altaffiltext{1}{Based on data obtained with the \textit{Hubble Space Telescope} operated by AURA, Inc. for NASA under contract NAS5-26555. }

\author{P. A. Oesch\altaffilmark{2,\dag},
R. J. Bouwens\altaffilmark{3}, 
G. D. Illingworth\altaffilmark{2}, 
I. Labb\'{e}\altaffilmark{3,4}, 
M. Trenti\altaffilmark{5}, 
V. Gonzalez\altaffilmark{2},
C. M. Carollo\altaffilmark{6}, 
M. Franx\altaffilmark{3}, 
P. G. van Dokkum\altaffilmark{7},
D. Magee\altaffilmark{2}
}

\altaffiltext{2}{UCO/Lick Observatory, University of California, Santa Cruz, CA 95064; poesch@ucolick.org}
\altaffiltext{\dag}{Hubble Fellow}
\altaffiltext{3}{Leiden Observatory, Leiden University, NL-2300 RA Leiden, Netherlands}
\altaffiltext{4}{Carnegie Observatories, Pasadena, CA 91101}
\altaffiltext{5}{University of Colorado, Center for Astrophysics and Space Astronomy,
389-UCB, Boulder, CO 80309, USA}
\altaffiltext{6}{Institute for Astronomy, ETH Zurich, 8092 Zurich, Switzerland}
\altaffiltext{7}{Department of Astronomy, Yale University, New Haven, CT 06520}

\begin{abstract}
We search for $z\sim10$ galaxies over $\sim$160 arcmin$^2$ of WFC3/IR data in the Chandra Deep Field South, using the public HUDF09, ERS, and CANDELS surveys, that reach to 5$\sigma$ depths ranging from 26.9 to 29.4 in $H_{160}$ AB mag. $z\gtrsim9.5$ galaxy candidates are identified via $J_{125}-H_{160}>1.2$ colors and non-detections in any band blueward of $J_{125}$. Spitzer IRAC photometry is key for separating the genuine high-z candidates from intermediate redshift ($z\sim2-4$) galaxies with evolved or heavily dust obscured stellar populations. After removing 16  sources of intermediate brightness ($H_{160}\sim24-26$ mag) with strong IRAC detections, we only find one plausible $z\sim10$ galaxy candidate in the whole data set, previously reported in Bouwens et al. (2011). The newer data cover a $3\times$ larger area and provide much stronger constraints on the evolution of the UV luminosity function (LF). If the evolution of the $z\sim4-8$ LFs is extrapolated to $z\sim10$, six $z\sim10$ galaxies are expected in our data. The detection of only one source suggests that the UV LF evolves at an accelerated rate before $z\sim8$. The luminosity density is found to increase by more than an order of magnitude in only $170$ Myr from $z\sim10$ to $z\sim8$. This increase is $\geq4\times$ larger than expected from the lower redshift extrapolation of the UV LF. We are thus likely witnessing the first rapid build-up of galaxies in the heart of cosmic reionization. Future deep HST WFC3/IR data, reaching to well beyond 29 mag, can enable a more robust quantification of the accelerated evolution around $z\sim10$.

\end{abstract}


\section{Introduction}

In recent years, great progress has been made in quantifying the evolution of the galaxy population at the end of cosmic reionization around $z\sim6$. Deep Hubble Space Telescope (HST) legacy fields, such as the Hubble Ultra-Deep Field \citep[HUDF;][]{Beckwith06} or GOODS \citep{Giavalisco04b}, and wide area ground-based imaging, have made it possible to study the evolution of the UV luminosity function (LF) across $z\sim4-6$ to great accuracy \citep[e.g.][]{Bouwens07,Oesch07,Iwata07,Ouchi04,Sawicki06,McLure09}.

Over the last two years, with the installation of the Wide Field Camera 3 (WFC3) onboard the HST, the observational frontier of galaxies has now been pushed into the reionization epoch, as deep WFC3/IR data led to the identification of more than 100 galaxy candidates at $z\sim6.5-8.5$ \citep[e.g.][]{Bouwens10c,Bouwens10a,Oesch10a,McLure10,Bunker10,Finkelstein10,Yan10,Wilkins10,Wilkins11,Lorenzoni11,Trenti11}. This is essential for estimating the contribution of galaxies to cosmic reionization. One of the most important conclusions from these studies is thus the realization that the UV luminosity density (LD) emitted by the galaxy population gradually falls towards higher redshifts. For example, the LD of the $z\sim3$ galaxy population is about an order of magnitude larger than that of the $z\sim8$ population, about 1.5 Gyr earlier.

How this evolves to even higher redshifts is still very unclear. A sizable galaxy population at $z\gtrsim9$ is expected based on the first estimates of stellar population ages of $z\sim7-8$ galaxies, indicating that these sources very likely started forming stars already at $z\gtrsim10-12$ \citep[e.g.][]{Labbe10a,Labbe10b,Finkelstein10,Gonzalez11}. This is still somewhat uncertain due to possible nebular line emission contaminating the Spitzer photometry \citep[e.g.][]{Schaerer10}. Nonetheless, an early epoch of star-formation is also required by the mean redshift of reionization as measured by WMAP \citep[$z_r=10.6\pm1.2$;][]{Komatsu11}, if galaxies are assumed to be the main drivers for this process.

However, previous searches for $z\gtrsim9$ sources in the pre-WFC3 era only resulted in very small samples of relatively low reliability candidates, none of which have been confirmed \citep[e.g.][]{Bouwens05,Stark07,Henry08,Henry09,Richard08}. This is mainly due to the extreme faintness of the $z\sim10$ galaxy population. Not only are these galaxies fainter due to their increased distance, but also they are expected at intrinsically lower luminosities. Additionally, the detection of such high-redshift sources is further complicated by the fact that they are invisible in optical data. Due to the highly neutral inter-galactic medium before the end of cosmic reionization, their UV photons are absorbed shortward of the redshifted \lya\ line, which shifts to $>1\micron$ at $z\gtrsim7$. Thus, these sources can only be seen in the NIR, where previous detectors were significantly lagging behind optical technology.

With 40$\times$ higher efficiency relative to NICMOS to detect high redshift galaxies in the NIR, WFC3/IR has the potential to change this and to push galaxy studies to beyond $z\sim9$. Several deep and wide area WFC3/IR data sets have been taken already and several more are upcoming. The challenge for identifying genuine $z>9$ sources in these data sets, is that these galaxies will only be visible in one band ($H_{160}$). This has already led to some controversy in the first searches for $z\sim10$ sources in the first-epoch WFC3/IR data over the Hubble Ultra Deep Field \citep[see e.g.][]{Bouwens11,Yan10}. In our recent analysis, which includes the full two-year WFC3/IR data over the HUDF as well as the shallower, but wider Early Release Science (ERS) data, \citet{Bouwens11} found only one single galaxy candidate detected at $>5 \sigma$ with an estimated redshift at $z\sim10.3$. Given that about three should have been detected, if the evolution of the LF continued as extrapolated from the trends established across $z\sim4$ to $z\sim6$, this provided first tentative evidence for an accelerated evolution in the galaxy population from $z\sim8$ to $z\sim10$. 

In this paper we significantly expand on our first $z\sim10$ analysis from WFC3/IR data presented in \citet{Bouwens11} by extending the search to all the deep WFC3/IR fields in the Chandra Deep-Field South area that have since become available.  The inclusion of the two deep HUDF09 parallel fields
is especially useful, since both reach just $\sim0.5$ mag shallower than the ultra-deep HUDF field but triple the search area for $\sim28-29$ AB mag sources. 
Additionally, we use different analysis tools developed by the first author that provide an independent analysis of the HUDF and ERS data.  While the Lyman-Break approach is similar in principle to that of \citet{Bouwens11} the use of independently tested software and procedures for the source detection and its analysis provides confirmation and validation.
The expanded data set also covers $>3\times$ the area at moderately deep $\sim26.5$ AB mag, thanks to the inclusion of the first epochs of CANDELS data over these fields.
This will be used to constrain the evolution of the galaxy population over the $\sim200$ Myr from $z\sim10$ when the universe was $\sim500$ Myr old to $z\sim8$ at $\sim700$ Myr. 

We start by describing the full data set in \S \ref{sec:data} and present the $z \sim10$ candidate selection and its efficiency in \S \ref{sec:selection}. In \S \ref{sec:LF} we present our new constraints on the LF at $z\sim10$. We will refer to the HST filters F435W, F606W, F775W, F850LP, F098M, F105W, F125W, F160W as \bFilter, \vFilter, \iFilter, \zFilter, \ymFilter, \yFilter, \jFilter, \hFilter, respectively.
Throughout this paper, we adopt $\Omega_M=0.3, \Omega_\Lambda=0.7, H_0=70$ kms$^{-1}$Mpc$^{-1}$, i.e. $h=0.7$. Magnitudes are given in the AB system \citep{Oke83}.

\begin{figure}[tbp]
	\centering
	\includegraphics[width=\linewidth]{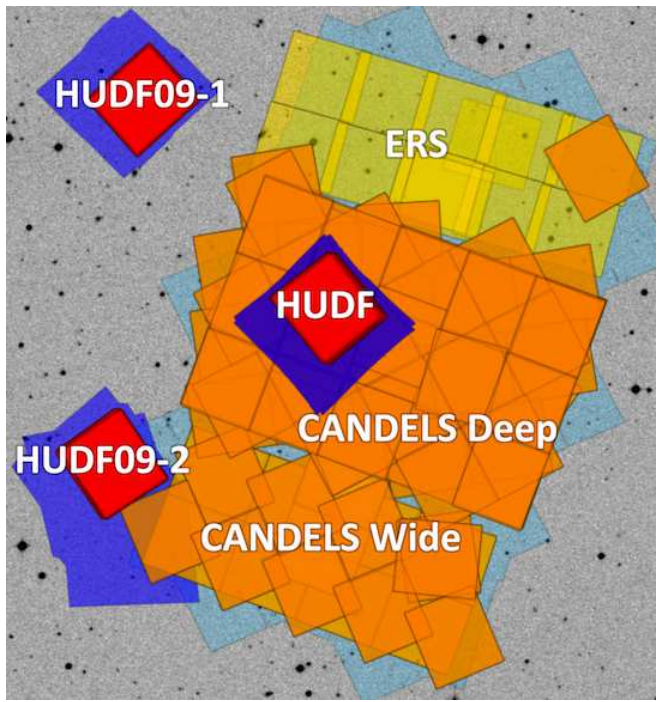}
  \caption{The WFC3/IR fields (orange) around the GOODS South area used in this analysis (see \S \ref{sec:data} and Table \ref{tab:data}). The HUDF/HUDF09-1/HUDF09-2 fields comprise the deepest NIR data available to date, reaching to $\gtrsim29$ AB mag. These three WFC3/IR pointings are covered in a total of 192 HST orbits, which obtained photometry in $Y_{105}$, $J_{125}$, and $H_{160}$. The ERS field consists of 60 orbits of WFC3/IR imaging in $Y_{098}$,  $J_{125}$, and $H_{160}$ spread over a 2$\times$5 tile. The CANDELS program is an on-going multi-cycle treasury survey consisting of two parts: CANDELS-Deep ($\sim60$ arcmin$^2$, limited at $H_{160,AB}\sim27.5$) and CANDELS-Wide ($\sim40$ arcmin$^2$, limited at $H_{160,AB}\sim26.9$). Only $J_{125}$ and $H_{160}$ data obtained before August 6, 2011 is used in this analysis (together with the shallow $Y_{105}$ imaging of CANDELS Wide).  }
	\label{fig:fields}
\end{figure}

\begin{deluxetable*}{lcccccccccc}
\tablecaption{Summary of 5$\sigma$ Depths\tablenotemark{a} of Observational Data Used in our Analysis\label{tab:data}}
\tablewidth{0 pt}
\tablecolumns{0}
\tablehead{\colhead{Field} & Area [arcmin$^2$] & \bFilter &\vFilter &\colhead{\iFilter} &  \colhead{\zFilter}  &\colhead{$Y_{098/105}$} & \colhead{\jFilter} & \colhead{\hFilter}  }
\startdata
HUDF09 & 4.7 & 29.2 & 29.6 & 29.4 & 28.8 & 29.1 & 29.3 & 29.4 \\ 
HUDF09-1\tablenotemark{b} & 4.7 & -- & 28.9 & 28.7 & 28.6 & 28.6 & 28.8 & 28.6 \\ 
HUDF09-2\tablenotemark{b} & 4.7 & 28.8 & 29.3 & 28.9 & 28.7 & 28.6 & 28.9 & 28.9 \\ 

ERS          & 41.3 & 27.8 & 28.0 & 27.5 & 27.2 & 27.4 & 27.8 & 27.6 \\ 
CANDELS-Deep\tablenotemark{b} & 63.1 & 27.8 & 28.0 & 27.5 & 27.2  & --  & 27.7  & 27.5 \\
CANDELS-Wide\tablenotemark{b} & 41.9 & 27.8 & 28.0 & 27.5 & 27.2  & 27.1  & 27.2  & 26.9 

\enddata

\tablenotetext{a}{Measured in circular apertures of 0\farcs25 radius.}
\tablenotetext{b}{New relative to \citet{Bouwens11} for $z\sim10$ search.}

\end{deluxetable*}

\section{The Data}
\label{sec:data}

Our analysis is based on the public WFC3/IR data sets that are available over the GOODS South fields \citep{Giavalisco04b} as a result of three different programs (HUDF09, ERS, and CANDELS), which we describe below. The outline of all fields is shown in Figure \ref{fig:fields}. The key feature of these fields is that they have WFC3/IR coverage in $J_{125}$ and $H_{160}$, which will be used to select $z\sim10$ sources. Additionally, they have  deep optical coverage as well as deep IRAC imaging, which is essential to exclude low redshift contamination (see section \ref{sec:selection}). 

All the WFC3/IR data has been reduced following standard procedures outlined, e.g., in \citet{Bouwens10c}. In particular, our reduction pipeline includes the subtraction of a super median image, careful image registration to the ACS frames and an automatic elimination of pixels affected by persistance. The final pixel scale of the images in our analysis is set to $0\farcs06$. A summary of the HST data used in this analysis can be found in Table \ref{tab:data}. The final resolution of the WFC3/IR data is $\sim0\farcs16$ (FWHM), and $\sim0\farcs09$ in the optical ACS data.

The ACS data on the GOODS fields we used are the v2 reductions that are publicly available from MAST\footnote{http://archive.stsci.edu/prepds/goods/} (M. Giavalisco and the GOODS Team, in preparation). The Spitzer data are the GOODS IRAC images \citep{Dickinson03} as made publicly available by the SIMPLE team \citep[see e.g.][]{Damen09}. 
Additionally, we include the newly acquired Spitzer IRAC [3.6] and [4.5] data over the HUDF field from the IUDF10 survey (proposal 70145, PI: Labb\'{e}). This so far adds $\sim130$h of observations, which increases the depth in both filters by an additional $\sim$0.4 mag.
Where available, we also matched our WFC3/IR sources with the publicly available GOODS-MUSIC multi-band photometry catalog of \citet{Santini09}.

\subsection{Full HUDF09 Data Set}
The HUDF09 program \citep[PI: Illingworth;][]{Bouwens10c} consists of 192 HST orbits to provide ultra-deep WFC3/IR imaging over three pointings centered on the HUDF \citep{Beckwith06} and its two parallel fields from the UDF05 program \citep[PI: Stiavelli;][]{Oesch07}. 
The program has been completed, providing the deepest IR images ever taken. It comprises $3\times4.7$ arcmin$^2$ imaging in the three filters $Y_{105}$, $J_{125}$, and $H_{160}$, reaching down to $H_{160,AB}=29.4, 28.6$, and 28.9 ($5\sigma$ in 0\farcs5 diameter apertures) for the HUDF, HUDF09-1, and HUDF09-2, respectively (see Fig. \ref{fig:fields}). For a more detailed description of this data set and the data reduction see \citet{Bouwens10c}.

\subsection{ Wide Area Data}
In addition to the ultra-deep HUDF09 data, we also analyzed shallower, wider area WFC3/IR imaging from the ERS and CANDELS programs, in order to constrain the volume density of more luminous star-forming galaxies at $z\sim10$.

The Early Release Science Data (ERS) provide WFC3/IR imaging of $\sim41$ arcmin$^2$ of the northern part of the GOODS South field. Two orbits of WFC3/IR imaging were obtained in each of the filters \ymFilter, \jFilter, and \hFilter, over a $2\times5$ grid of pointings (60 orbits in total). These data are reduced in an analogous way to our HUDF09 data, and are aligned and drizzled to the GOODS ACS mosaics after rebinning to a $0\farcs06$ pixelscale. These data reach to $H_{160}=27.6$ \citep[see also][]{Bouwens10c}. For a more detailed description of this data set see \citet{Windhorst11}.

The last two fields included in our analysis are obtained as part of the Multi-Cycle Treasury program CANDELS \citep[PI: Faber/Ferguson;][]{Grogin11,Koekemoer11}. In particular, we include the first six visits of the CANDELS-Deep program (obtained until August 6, 2011), which covers the central part of GOODS South in $3\times5$ tiles with $\sim6000$ s exposures in both \jFilter\ and \hFilter\ in a total of 92 orbits. This data covers $\sim63$ arcmin$^2$ and reaches to $H_{160,AB}\sim27.5$ mag. 
Additionally, we also included the imaging data of the supernova follow-up program of CANDELS (PI: Riess), which adds imaging over two pointings over CANDELS Deep (one of which is essentially centered on the HUDF).
Finally, we made use of the 29 orbits of WFC3/IR data of the CANDELS-Wide survey ($Y_{105}$, $J_{125}$ and $H_{160}$, obtained until March 29, 2011). These comprise 9 WFC3/IR pointings ($\sim42$ arcmin$^2$), completing the coverage of the GOODS South field, and reach to $H_{160,AB}=26.9$ ($\sim2000$ s exposures).
As for the ERS, the WFC3/IR data of the CANDELS program has been aligned to the GOODS ACS mosaics with a pixel scale of $0\farcs06$. The part of the CANDELS field overlapping with the WFC3/IR HUDF has been omitted when analyzing this data set in order not to duplicate the analysis of that area.

We will subsequently  refer to the combination of the ERS and the two CANDELS fields as `Wide Fields'.

\section{Source Selection}
\label{sec:selection}

\subsection{Catalog Construction} 
Source catalogs are derived with the SExtractor program \citep{Bertin96}, which is used to detect galaxies in the $H_{160}$ images and perform matched aperture photometry on PSF-matched images. The colors used here are based on isophotal apertures derived from the $H_{160}$ images, and total magnitudes are measured in standard 2.5 Kron apertures, corrected by 0.2 mag in order to account for flux loss in the PSF wings. 

The detection significance of sources was established in $0\farcs25$ radius apertures. The RMS maps were scaled based on the detected variance in 1000 random apertures for each WFC3/IR frame on empty sky regions after  $3\sigma$ clipping. This procedure ensures that the SExtractor weight maps correctly reproduce the actual noise in the images. Subsequently only sources with signal-to-noise ratios larger than 5 (in $0\farcs25$ radius apertures) in $H_{160}$ are considered.

\begin{figure}[tbp]
	\centering
	\includegraphics[width=\linewidth]{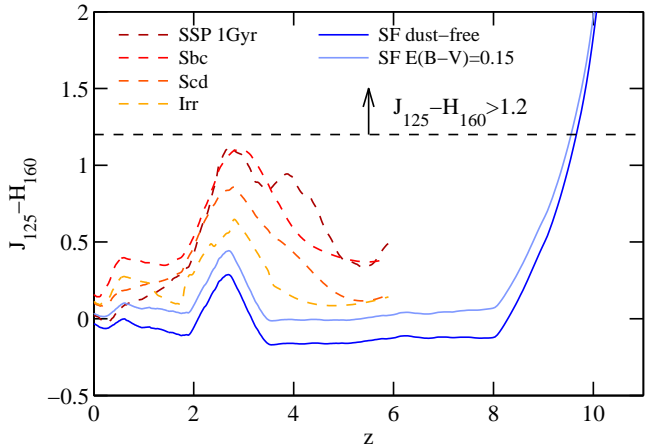}
	
  \caption{Color selection of $z>9.5$ galaxies (\S \ref{sec:candsel}). Shown are the $J_{125}-H_{160}$ colors of different types of galaxies as a function of their redshift. Star-forming galaxies are shown as solid blue lines. The lighter blue spectral energy distribution (SED) is reddened by E(B-V)$=0.15$ mag using a \citet{Calzetti00} dust law. As the Ly$\alpha$ absorption due to inter-galactic neutral hydrogen shifts into the $J_{125}$ band, galaxies start to exhibit progressively redder colors beyond $z\sim8$. A selection with $J_{125}-H_{160}>1.2$ thus identifies galaxies at $z\gtrsim9.5$.  Strong Balmer and 4000 \AA\ breaks in evolved $z\sim3$ galaxies, combined with some dust obscuration can also result in very red $J_{125}-H_{160}$ colors. The dashed lines correspond to more evolved galaxies dominated by progressively older stellar populations from the library of \citet{Coleman80} as well as a 1 Gyr old single stellar population (SSP) from the library of \citet{Bruzual03}.   Spitzer IRAC data provides a way to separate these different populations at intermediate and very high redshifts (see text and Fig. \ref{fig:maglim}). }
	\label{fig:colsel}
\end{figure}

\begin{figure}[tbp]
	\centering
	\includegraphics[width=\linewidth]{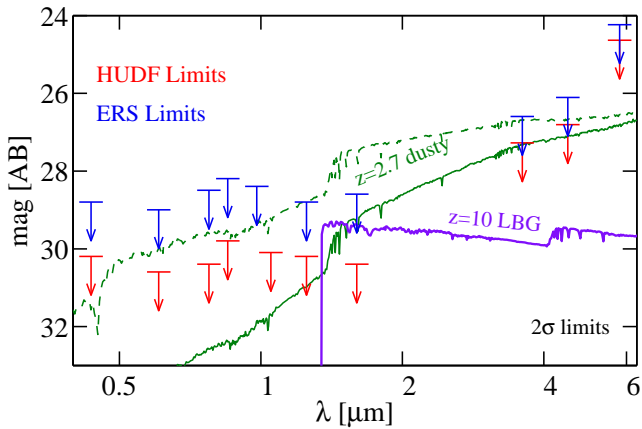}
  \caption{Sensitivity limits ($2\sigma$) of some of the data used in this analysis. The red arrows indicate the limits of the ACS, WFC3/IR and IRAC data over HUDF, while blue arrows correspond to the ERS data set. The SED of a $z=10$ star-forming galaxy is plotted in purple, showing the distinguishing feature of complete absorption shortward of the redshifted Ly$\alpha$ line. Evolved and dusty galaxy SEDs can mimic the same features as a $z\sim10$ source, i.e. red $J_{125}-H_{160}$ and undetected in the optical data, unless this is extremely deep. Such sources are typically detected in the IRAC data. Alternatively, the solid and dashed green lines show possible SEDs of $z\sim2.7$ galaxies which could escape the IRAC detection at $2\sigma$. These correspond to evolved galaxies (500 Myr), with moderate amounts of dust extinction, and are normalized to the 5 $\sigma$ detection limits of the $H_{160}$ bands. Note that much larger amounts of dust can be hidden for interloper galaxies at the detection limit of the HUDF (green solid, $A_V=2.0$ mag) than for the ERS (green dashed, $A_V=0.6$ mag). Thus even with IRAC it can be a challenge to separate real $z\sim9-10$ sources from lower redshift contaminants. }
	\label{fig:maglim}
\end{figure}

\subsection{\Jdrop\ Candidate Selection}
\label{sec:candsel}
Galaxies at $z>9.5$ are expected to exhibit very red $J_{125}-H_{160}$ colors since the redshifted, strong \lya\ absorption (by the predominantly neutral inter-galactic hydrogen) cuts into the flux in the $J_{125}$ filter (see Figure \ref{fig:colsel}). This makes such high redshift galaxies completely invisible blueward of $J_{125}$, and we use this fact for their identification.

As can be seen from Figure \ref{fig:colsel}, however, also passively evolving or dusty galaxies at intermediate redshifts ($z\sim2.5-4$) can exhibit similarly red colors in $J_{125}-H_{160}$. While the requirement of optical non-detections removes the bulk of lower redshift contamination, certain intermediate redshift galaxies with evolved or dusty stellar populations can still be included due to the fact that the optical data does not reach deep enough, if at similar depth as the IR (see Fig. \ref{fig:maglim}). 
Deep Spitzer IRAC data provides a way to identify contaminating galaxies. These are expected to exhibit very red $H_{160}-[3.6\micron]$ colors, which discriminates them from genuine $z\sim10$ candidates.

To exclude possible low-redshift contamination, we thus use two steps to select $z\gtrsim9.5$ galaxy candidates. For the first step, the primary criteria are based on HST data only:
\[
	(J_{125}-H_{160})>1.2
\]
\[
	S/N(H_{160}) > 5 \quad \wedge \quad  S/N(<J_{125})<2 \quad \wedge \quad \chi^2_{opt}<\chi^2_{cut}
\]
Additional to excluding objects that are detected in any band blueward of $J_{125}$ at more than $2\sigma$, we include a cut in the optical $\chi^2_{opt}$ value of a galaxy \citep[see e.g.][]{Bouwens10c,Bouwens11}. This is computed from the $0\farcs25$ radius aperture fluxes as $\chi^2_{opt} = \sum_i\mathrm{sign}(f_i)\left(f_i/\sigma_i\right)^2$, where the sum runs over all the bands available in the given data set blueward of $J_{125}$, i.e. it includes all the available optical data as well as the NIR band $Y_{105}$ for the HUDF09 data, and $Y_{098}$ in the ERS. The relatively large apertures were chosen in order sample $>70\%$ of the light of point-like sources.

The limiting $\chi^2_{cut}$ are derived from photometric scatter simulations. They are set to exclude the majority of interlopers which remain undetected at $2\sigma$ purely due to photometric noise, but not to cut a substantial fraction of galaxies with real zero flux in the optical bands. 
The scatter simulations utilize all galaxies in our catalogs that are $1-3$ mag above the completeness limit, applying photometric Gaussian noise from 1 mag fainter sources. From these simulations it is clear that contamination is mainly an issue at 0.75 mag above the completeness limits, but that $\sim60-80$\% of contaminants can be eliminated by using a $\chi^2_{opt}$ limit of $\chi^2_{cut}=2.8$ or 2.4, for 5 filters or 4 filters, respectively. 
In the HUDF09 data, the resulting number of expected contaminants due to photometric scatter is thus reduced from $\sim0.5$ source per WFC3/IR field to $\sim0.1$ source.

On the other hand, the adopted $\chi^2_{cut}$ limits do remove an additional $\sim20$\% of sources with real zero flux, simply due to Gaussian statistics. This reduction of the real galaxy sample is reflected in our subsequent analysis in the reduction of the selection volume.

\begin{deluxetable*}{lcccccccccc}
\tablecaption{The $z\sim10$ Galaxy Candidate\label{tab:phot}}
\tablewidth{0 pt}
\tablecolumns{7}
\tablehead{\colhead{ID} & $\alpha$ & $\delta$ &\colhead{$H_{160}$}  &\colhead{$J_{125}-H_{160}$} &  \colhead{S/N$_{H160}$}   }

\startdata

HUDFj-39546284 &  03:32:39.54 &  -27:46:28.4 &  $28.8\pm0.2$ & $>2.02$ &  6.3

\enddata

\end{deluxetable*}

\begin{figure*}[tbp]
	\centering
	\includegraphics[scale=0.75]{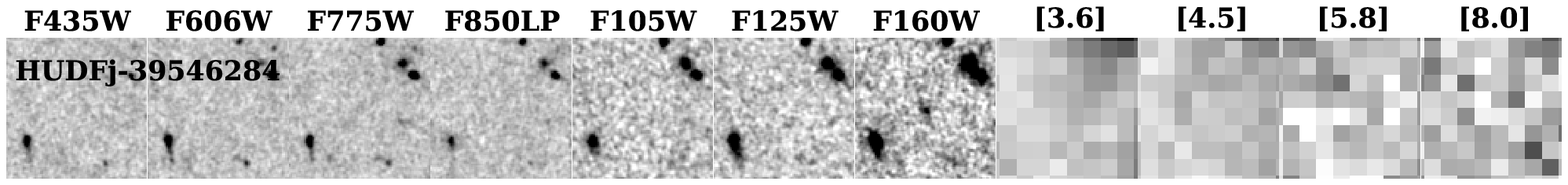}
  \caption{5\arcsec$\times$5\arcsec images of our only viable $z\sim10$ galaxy candidate \citep[see also][]{Bouwens11}. The images show, from left to right, \bFilter, \vFilter, \iFilter, \zFilter, \yFilter, \jFilter, \hFilter, IRAC [3.6], [4.5], [5.8], and [8.0]. The \jFilter\ and \hFilter\ images shown here are combined from both the HUDF09 and the CANDELS survey data (which add 0.05 mag in depth).
The IRAC [3.6] and [4.5] images include the additional $\sim130$h of data obtained over the HUDF as part of the IUDF10 program so far.
  As can be seen, the source is only significantly detected in $H_{160}$. The IRAC [$3.6\micron$] data shows a non-negligible excess of flux in the vicinity of the candidate. However, this flux is very likely to be associated with the brighter neighboring source to the upper right.  After subtraction of all nearby sources in IRAC, the candidate is undetected in both IRAC bands at less than $1\sigma$. The images are oriented N top, E left.}
	\label{fig:stampz10}
\end{figure*}

All galaxies passing the above selection criteria,  using both the ACS and WFC3/IR data, are retained and analyzed individually. These total to 17 sources with $H_{160,AB}$ in the range $23.6-28.8$ mag; one source in the HUDF, none in the parallel HUDF09 fields, three in the ERS and 8 and 5 in the CANDELS Deep and Wide, respectively (see Tables \ref{tab:phot} and \ref{tab:photContamin}).

Interestingly, only one source (the previously reported galaxy with $H_{160}\sim29$ mag from Bouwens et al. 2011) did pass our selection in the three deep HUDF09 fields, while the shallower CANDELS and ERS fields contribute a total of 16 sources (all with $H_{160,AB}\lesssim26$ mag). Upon inspection of their images, it turns out that all these brighter sources are very well detected in the IRAC data, even in the shallow 8.0$\micron$ band. 
Their measured $H_{160}-[3.6]$ colors are in the range $1.6-4.3$, which, for a $z\sim10$ source, would correspond to a UV continuum slope  $\beta\gtrsim-0.2$, or a dust reddening with $A_V>1.6$ mag. Given that galaxies at the bright end of the $z\sim7$ population are measured to have very low extinction values and continuum slopes of $\beta\simeq-2.0\pm0.2$ \citep[see e.g.][]{Bouwens10b,Finkelstein10,Dunlop11,Wilkins11b}, the extremely red colors of these galaxies rules out $z\gtrsim9$ solutions with any sensible SED.

All sources with IRAC detections are thus removed from our sample of potential $z\sim10$ galaxies, reducing the sample to one single candidate in the HUDF,  previously reported in \citet[][see Figure \ref{fig:stampz10} and Table \ref{tab:phot} ]{Bouwens11}. The 16 removed sources are shown in the appendix in Figure \ref{fig:stampsContamin} and listed in Table \ref{tab:photContamin}.

\subsection{The $z\sim10$ Galaxy Candidate}

Images of the only possible $z\sim10$ galaxy candidate are shown in Figure \ref{fig:stampz10}. The source is detected at 6.3 $\sigma$ in $H_{160}$ (measured in circular apertures of $0\farcs25$ radius). This is higher but completely consistent with the \citet{Bouwens11} significance estimates, which are based on smaller apertures. 
As can be seen, the source is not significantly detected in any other band. Its value of $\chi^2_{opt}=2.77$ is very close, but just below the limit of $\chi^2_{cut}=2.8$. This is mainly due to a 1.5$\sigma$ flux excess in $i_{775}$, which appears to be due to an extended structure in the background of that image. When adopting smaller apertures, the $\chi^2_{opt}$ value is found to be reduced, indicating also that this excess of flux is not associated with the source itself.

The Spitzer IRAC 3.6$\mu$m data shows some flux from a nearby source. However, after subtraction of all the neighboring sources in IRAC, the candidate is undetected ($0.09\sigma$), with a $2\sigma$ upper limit on its IRAC [3.6] magnitude of $>27.2$ mag AB (Gonzalez et al., in prep.). This thus corresponds to $H_{160}-[3.6]<1.6$ at $2\sigma$, which is much bluer (by $>0.4$ mag) than the typical low-redshift contaminants that we culled from our sample (see also next section). 
After adding the newly acquired IRAC data from the IUDF10 program (Spitzer proposal 70145, PI: Labbe) to the GOODS IRAC data and removing neighboring sources, the $z\sim10$ candidate is also undetected at [4.5] ($S/N=0.3$) providing added weight to the likelihood of it being at high rather than low reshift.

\begin{figure}[tbp]
	\centering
	\includegraphics[width=\linewidth]{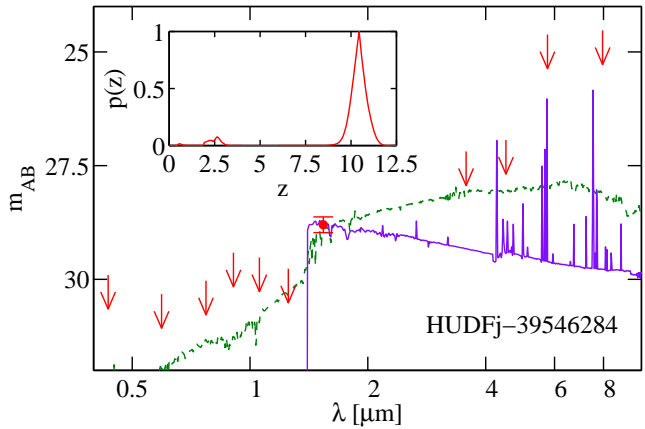}
  \caption{Best-fit SED of the only viable $z\sim10$ candidate, previously reported in \citet{Bouwens11}. The measured photometry is shown with the red circle with errorbars and with 2$\sigma$ upper limits in case of non-detections. All fluxes and flux errors were used in the SED fit, however, even if they were negative. The best fit SED is found at $z_{phot}=10.4$ and is shown as a solid line ($\chi^2_{best} = 7.0$), corresponding to a dust-free, young galaxy. The best template for a low redshift solution is also shown  as the dashed green line. This SED has a redshift of $z_{lowz}=2.7$, and is passive with an age of 650 Myr and of rather low mass (only $3\times10^{8}$ \msol). Additionally, the low-z SED is reddened with $E(B-V)=0.1$ mag. Based on its larger $\chi^2$ value ($\chi^2_{lowz}=13.7$), this SED is formally excluded at $>90\%$ probability, however. The redshift probability function is shown in the inset in the upper left. The sharp decrease above $z\sim10.5$ of $p(z)$ is due to the use of a LF prior at $z>8$ and due to the fact that \lya\ absorption starts affecting the $H_{160}$ band (which would require the source to be brighter intrinsically).}
	\label{fig:JdropSED}
\end{figure}

We derive the photometric redshift of the candidate using the code ZEBRA \citep{Feldmann06,Oesch10c} with synthetic stellar population models from \citet{Bruzual03} to which we added nebular continuum and line emission following, e.g., \citet{Schaerer09}.
Using the full 11 band fluxes and flux errors, we derive a photometric redshift for this source of $z_{phot} = 10.4^{+0.5}_{-0.4}$, with a likelihood of a low redshift solution at $z<8$ of $<6\%$. The full spectral energy distribution (SED) of the source and its redshift likelihood function are shown in Figure \ref{fig:JdropSED}.  The best-fit SED corresponds to a very young, dust-free star-burst (see also Gonzalez et al. in prep.).

The best low redshift solution is found at $z_{lowz}=2.7$, for an evolved, very low mass galaxy SED ($M=3\times10^8$ \msol) with moderate extinction. Interestingly, this SED is expected to be  detected only at $\sim1.5\sigma$ in $J_{125}$, but it nevertheless has a significantly higher $\chi^2$ value (13.7 compared to $\chi^2_{best}=7.0$). 
Deeper HST data shortward of the break would be extremely useful in order to constrain the possible non-detection of the source shortward of 1.4 \micron. As with other high-redshift catalogs, the added shorter-wavelength optical/near-IR data would play a key role in helping to further tighten its photometric redshift measurement.

\subsection{The Dusty and Evolved Contaminants of $J_{125}$-Dropout Selections}
\label{sec:contaminants}

The properties of the 16 intermediate brightness sources that did not pass the IRAC non-detection criteria are discussed in the appendix. From fitting their SEDs, these sources are found to be mostly massive galaxies ($M>5\times10^{10}$ \msol) with obscured but evolved stellar populations at $z\sim2-4$.

Interestingly, all these galaxies are essentially limited to $H_{160}\sim24-26$ mag. This is $\sim$1 mag brighter than the detection limits of our bright survey fields (see Figure \ref{fig:ContaminMz} in the appendix), and thus suggests that such very red galaxies have a somewhat peaked luminosity function. This will have to be confirmed and quantified with future wide area data. However, at face value, it would appear that contamination from such sources with $J_{125}-H_{160}>1.2$, and with $H_{160}-[3.6]\gtrsim2$ is less problematic at fainter magnitudes. We note, however, that contamination from similar galaxies with less extreme colors (which may be more abundant also at $H_{160}>26$ mag) may still be non-negligible due to photometric scatter (see also \S \ref{sec:additionalcontamin}).

Furthermore, it is interesting to note that it is very difficult to construct clean $z\sim10$ galaxy samples purely based on HST data alone. Even if we were to increase the color criteria to $J_{125}-H_{160}>2.0$, there would still be two contaminating galaxies in the sample together with our only viable $z\sim10$ candidate. Thus, also for future constraints on the bright end of the $z\sim10$ LF, it will be important to perform additional follow-up studies to validate the candidates, e.g. with Spitzer. This will be less of a concern for future $z\sim9$ galaxy samples, which can be obtained, e.g.,  based on new F140W filter data. In such a data set intrinsically red galaxies can be sorted out by requiring a blue continuum across F140W and $H_{160}$.

\subsection{Possible Sources of Sample Contamination}
\label{sec:additionalcontamin}

Here, we only give a brief summary of the possible sample contamination. For a thorough discussion we refer the reader to \citet{Bouwens11}. 

Essentially, the only probable chance for contamination is due to photometric scatter of a red lower redshift source. 
We estimate this to be a 10\% chance based on our photometric scatter experiments described in section \ref{sec:candsel}, including the $\chi^2_{opt}$ cuts.

Other typical contaminants to LBG selections such as very cool dwarf stars and supernovae can essentially be excluded based on the relatively blue $J_{125}-H_{160}$ ($<1.1$) colors of stellar SEDs, on the fact that the source is detected in both the first and the second year of the HUDF09 WFC3/IR data, and due to the fact that the candidate shows clear signs of an extended morphology.

Additionally, it is very unlikely that this source is  spurious, since the flux distribution in circular apertures randomly distributed over empty regions of the HUDF $H_{160}$ image are nearly exactly Gaussian, and the source is well detected at $6.3\sigma$.

\section{The Abundance of $z\sim10$ Galaxies}
\label{sec:LF}
In this section we compute the expected abundance of $z\sim10$ galaxies in our dataset, and derive constraints on the $z\sim10$ LF based on our data.

\subsection{Completeness and Selection Functions}

 In order to estimate the number of sources we expect in our data from a given LF, we have to estimate the completeness as a function of magnitude $C(m)$ and selection function as a function of redshift and magnitude $S(z,m)$. Following \citet{Oesch07,Oesch09}, this is done by inserting artificial galaxies in the observational data and rerunning the source detection with the exact same setup as for the original catalogs. This is done for each of our fields individually. 

Two sets of simulations were run. In the first set, we follow \citet{Bouwens03}, where the artificial galaxies are `cloned' from the $z\sim4$ dropout sample of the GOODS and HUDF fields. The images of these sources are adjusted for surface brightness dimming, the difference in angular diameter distance, as well as a size scaling of $(1+z)^{-1}$ as observed for the Lyman Break galaxy population across $z\sim3-7$ \citep[see e.g.][]{Ferguson04,Bouwens04a,Oesch10b}. These are then inserted in the observed images with galaxy colors as expected for star-forming galaxies between $z=8$ and $z=12$. When computing the galaxy colors we assume a distribution of UV continuum slopes with $\beta=-2.5\pm0.4$ \citep[see e.g.][]{Bouwens09b,Bouwens10b,Finkelstein10,Stanway05}. 

From these simulations, we compute the completeness as a function of observed $H_{160,AB}$ magnitude for each field, taking into account the scatter and offsets between input and output magnitudes. Additionally, we compute the selection probabilities as a function of redshift and magnitude by measuring the fraction of sources that meet our selection criteria. By construction, galaxies are selected at $>50\%$ at redshifts $z\gtrsim9.5$.

In the second set of simulations, we repeat the above procedure. However, instead of using observed galaxy images, we use theoretical galaxy profiles from a mix of exponential (Sersic $n=1$) and de Vaucouleur (Sersic $n=4$) profiles. The size distribution is chosen to be log-normal, again with the same size scaling as a function of redshift. The completeness and selection functions of our two procedures are in excellent agreement. This demonstrates the reliability of our approach, which appears to be essentially independent of the adopted galaxy profiles \citep[unlike what has been claimed elsewhere, e.g.][but see also Bouwens et al. 2010b]{Grazian11}.

\begin{figure}[tbp]
	\centering
	\includegraphics[width=\linewidth]{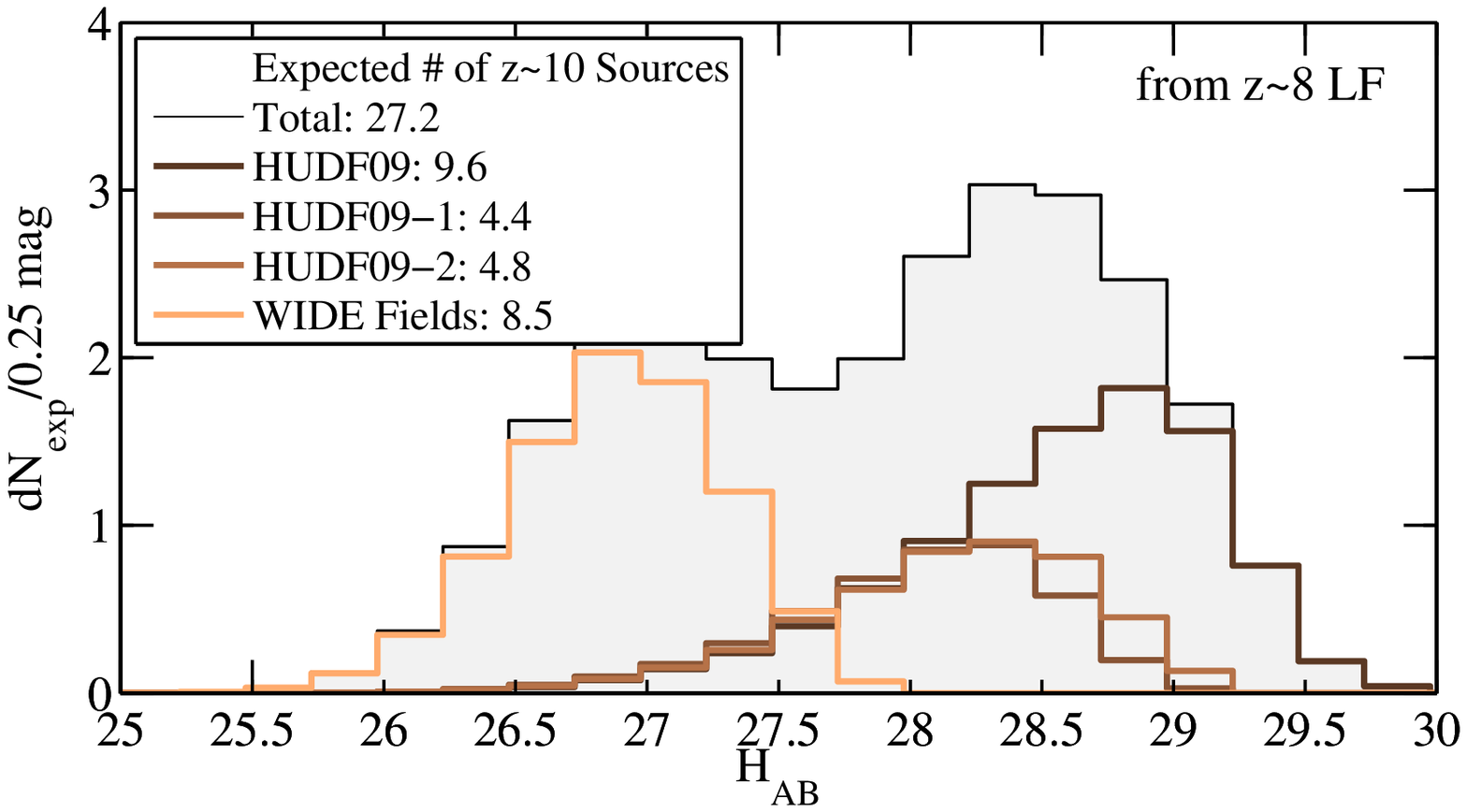}
	\includegraphics[width=\linewidth]{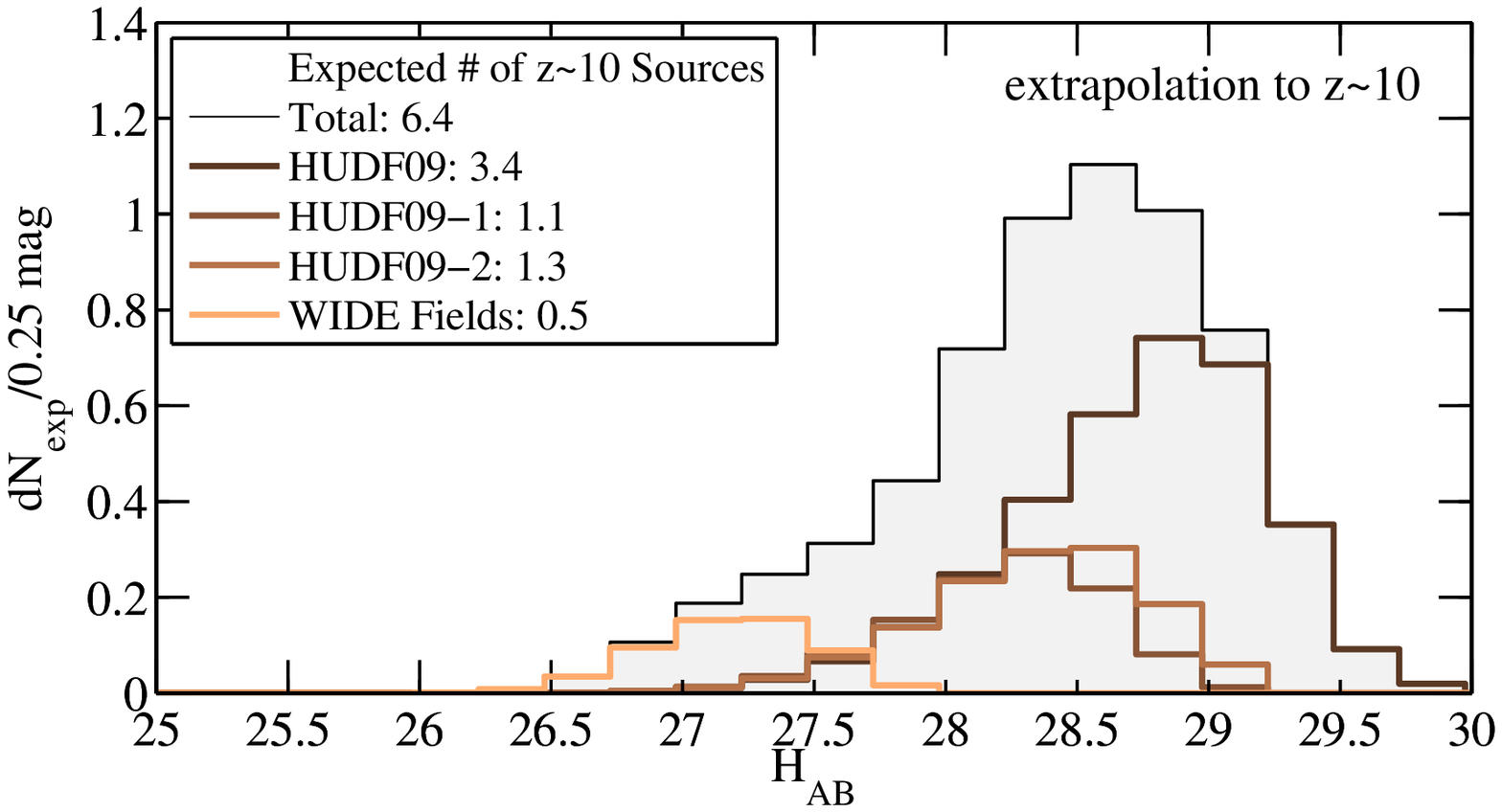}
  \caption{Expected number of $z\sim10$ candidates in the different fields assuming the LF to be as measured at $z\sim 8$ \citep[top;][]{Bouwens10c} or as expected from extrapolating the $z\sim4-8$ trends to $z\sim10$ (bottom). The one detected source is in stark contrast to the $\sim27$ sources expected if the LF was constant across the 170 Myr from $z\sim10$ to $z\sim8$. The LF appears to evolve at a significantly accelerated pace with respect to the empirical evolution observed across $z\sim4-8$. 6 sources at $z\sim10$ should be detected when extrapolating the lower redshift trends, while only one probable candidate is found.}
	\label{fig:Nexp}
\end{figure}

\subsection{The Expected Number of $z\sim10$ Sources}
\label{sec:Nexp}

The expected number of sources in a given magnitude bin $m_i$ can be estimated for any given LF, $\phi(M)$, through:
\[
N^\mathrm{exp}_i  = \int_{\Delta m} dm \int dz \frac{dV}{dz} S(m,z)C(m)\phi(M[m,z])
\]
The result of this calculation is shown in Figure \ref{fig:Nexp} for each individual field using the observed LFs at $z\sim8$ \citep{Bouwens10c}, as well as that expected at $z\sim10$. The latter is based on extrapolating the evolution of the Schechter function parameters presented in \citet{Bouwens10c}, which is based on fitting the likelihood contours of LBG LFs across $z\sim4-8$. In particular, significant evolution is only seen  in the characteristic luminosity $M_*$, which is found to dim by 0.33 mag per unit redshift. 
The other two Schechter function parameters were kept constant, in agreement with the small evolutionary trends that were found at low significance in \citet{Bouwens10c}. For reference, the final parameter evolution we use in this work is: 

\[M_*(z) = -21.02 ~\mathrm{mag} + (z-3.8) \times 0.33 ~\mathrm{mag}  \]
\[\phi_*  = 1.14\times10^{-3} ~\mathrm{Mpc}^{-3}\mathrm{mag}^{-1} =\mathrm{const}\]
\[\alpha  = -1.73 =\mathrm{const}\]

As can be seen in Figure \ref{fig:Nexp}, the wide-area, shallow data of CANDELS and ERS are very useful for constraining the evolution in the bright end of the LF. In particular, if there was no evolution in the LF from $z\sim10$ to $z\sim8$, the wide area data should contain 8.5 $z\sim10$ sources.
In total, we would expect to see $\sim27$ sources if the LF was unchanged over the 170 Myr from $z\sim8$ to $z\sim10$. This implies that the wide area data provides essentially 30\% of the total search power in case of an LF evolution in $\phi_*$ only.

Given that we only detect one probable candidate, the LF appears to drop faster than expected from the empirical lower redshift extrapolation. In particular, we do not detect the $\sim6$ galaxies that we would have expected to find at $z\sim10$ if the lower redshift trends remained valid. 
From these extrapolations we predicted to find three sources in the HUDF \citep[consistent with the expectations from][]{Bouwens11} and about one in each of the two HUDF09 parallel fields.

The Poissonian probability to find $\leq1$ source, given that 6 are expected is $<2\%$. This remains significant, even after including cosmic variance, which adds an additional uncertainty on these low number counts of about $45-50\%$  for an individual WFC3/IR pointing \citep[see e.g.,][]{Trenti08,Robertson10}. We derive an upper limit of 6\% to the probability of finding $\leq1$ source in our search area, based on the cosmic variance calculator of \citet{Trenti08} and combining the number counts uncertainty in the different fields assuming the final distribution is Gaussian (justified by the central limit theorem).
Therefore, the detection of an accelerated evolution relative to the low-redshift extrapolation is significant at $\geq94\%$.

\subsection{Constraints on the $z\sim10$ Luminosity Function}
\label{sec:LFconstraints}

\begin{figure}[tbp]
	\centering
	\includegraphics[width=\linewidth]{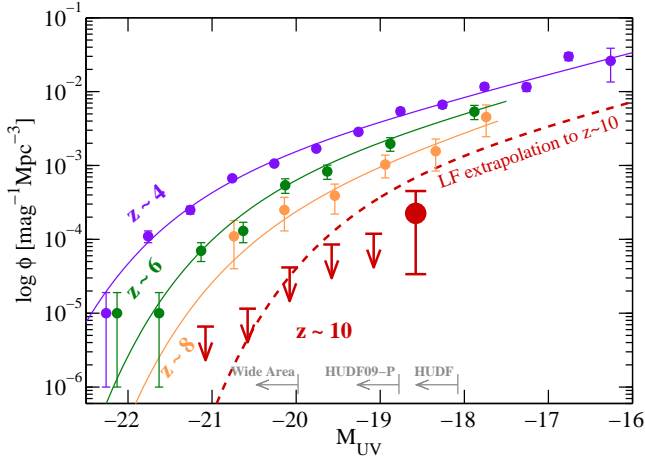}
  \caption{Constraints on the $z\sim10$ LF from our combined data set, evaluated in 0.5 mag bins. The upper limits correspond to $1\sigma$ Poissonian limits including the additional uncertainty of 50\% cosmic variance per pointing. It is clear that the luminosity function evolves strongly from $z\sim8$ to $z\sim10$, as our upper limits are a factor $\sim$2-5 below the measured $z\sim8$ LF of \citet[orange data;][]{Bouwens10c}.  The expected $z\sim10$ LF as extrapolated from fits to lower redshift LBG LFs is shown as a dashed red line. Using this LF, we would expect to detect six sources in the full data set ($\sim3.4$ in the HUDF, and $\sim1.2$ in each of the HUDF09 parallels; see Figure \ref{fig:Nexp}). For comparison also the $z\sim4$ and $z\sim6$ LFs are plotted \citep{Bouwens07}, showing the dramatic build-up of UV luminosity across $\sim1$ Gyr of cosmic time. The light gray vectors along the lower axis indicate the range of luminosities over which the different data sets dominate the $z\sim10$ LF constraints. }
	\label{fig:LFevol}
\end{figure}

The accelerated evolution in the UV LF can also be seen from our constraints on the $z\sim10$ LF. The stepwise LF is computed using an approximation of the effective selection volume as a function of observed magnitude $V_{\rm eff}(m) = \int_0^\infty dz \frac{dV}{dz} S(z,m)C(m)$. The LF in bins of absolute magnitude is then given by $\phi(M_i)dM = N^{\rm obs}_i/V_{\rm eff}(m_i)$. This is shown in Figure \ref{fig:LFevol}, where the LF was evaluated in bins of 0.5 mag, and non-detections correspond to $1\sigma$ upper limits including the effects of 50\% cosmic variance per pointing.

As can be seen in Figure \ref{fig:LFevol},  the current constraints on the $z\sim10$ LF are significant even at bright magnitudes where the wide area data are particularly valuable.
These data sets reduce the upper limits by more than an order of magnitude relative to using only the HUDF09 fields, therefore indicating that the $z\sim10$ LF at $M_{1400}<-20$ drops by a factor $\sim4-5$ with respect to the observed LF at $z\sim8$. However, from these shallow data sets, no constraints can be obtained on any accelerated evolution of the UV LF around $z\sim8-10$. This only becomes apparent at $M_{1400}>-20$ (corresponding to $H_{160,AB}\sim27.5$ mag). Such faint limits are only probed by the HUDF09 data set. In particular, at $M_{1400}=-19$, the upper limit on the LF is a factor $\sim3$ below the expectation (Fig. \ref{fig:LFevol}). Thus, it is clear that data reaching to deeper than $H_{160,AB}=28.5$ mag will be necessary to further constrain the drop in the LF in the future  (which is beyond the reach of the current MCT programs).

In order to quantify the change in the LF from $z\sim10$ to $z\sim8$ more robustly, we consider two possible scenarios. First, we assume the accelerated LF evolution occurs only in $M_*$, at a constant rate since $z\sim6$, and we fit the Poissonian likelihood for the observed number of sources. This can be written as $\cal{L} $ $= \prod_{j} \prod_i P(N^{\rm obs}_{j,i},N^{\rm exp}_{j,i})$, where $j$ runs over all fields, and $i$ runs over the different magnitude bins, and $P$ is the poissonian probability.

Our extrapolation of the UV LF is a modification of the fitting formulae of \citet{Bouwens10c}. We thus use $\phi_*(z) = 1.14\times10^{-3} $ Mpc$^{-3}=$const. and $\alpha(z)=-1.73=$const., and assume: $M_*(z)=-	20.29+\zeta(z-6)$. We then fit for $\zeta$, finding $\zeta = 0.58^{+0.14}_{-0.11}$, which results in an estimate for $M_*(z=10)=-18.0\pm0.5$ mag.

Alternatively, we assume $M_*(z) = -19.63=$const. (as derived for $z\sim8$ from our empirical extrapolation), $\alpha=-1.73$, and we fit only for an evolution in the normalization with redshift relative to the $z\sim8$ LF. This results in $\phi_*= 1.14\times10^{-3} 10^{-\Upsilon(z-8)} $ Mpc$^{-3}$, with best fit $\Upsilon=0.54^{+0.36}_{-0.19}$. Thus, using this extrapolation, the normalization of the UV LF from $z\sim10$ to $z\sim8$ is expected to increase by a factor 12. These results are summarized in Table \ref{tab:summary}.

\begin{figure}[tbp]
	\centering
	\includegraphics[width=\linewidth]{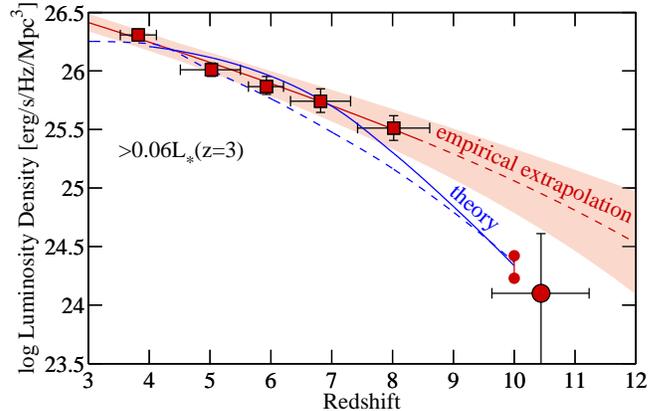}
  \caption{The evolution of the UV luminosity density $\rho_L$ above $M_{1400}=-18$ mag ($>0.06 L_*^{z=3}$). The filled circle at $z\sim10.4$ is the luminosity density directly measured for our only $z\sim10$ galaxy candidate. The two connected dots at $z=10$ show the range of possible LD values, given the two simple, accelerated extrapolations of the UV luminosity function described in section \ref{sec:LFconstraints}. The red line corresponds to the empirical LF evolution from \citet{Bouwens10c}. Its extrapolation to $z>8$ is shown as dashed red line. The $\rho_L$ data at $z\sim4-8$ is taken from \citet{Bouwens07,Bouwens10c}. As can be seen, $\rho_L$ increases by more than an order of magnitude in the 170 Myr from $z\sim10$ to $z\sim8$, indicating that the galaxy population at this luminosity range evolves by a factor $\geq4$ more than expected from low redshift extrapolations.
The predicted $\rho_L$ evolution of the semi-analytical model of \citet{Lacey11} is shown as dashed blue line, and the theoretical model prediction of \citet{Trenti10} is shown as blue solid line. These reproduce the expected luminosity density at $z\sim10$ remarkably well.
   }
	\label{fig:LDevol}
\end{figure}

\begin{deluxetable*}{ccccccc}
\tablecaption{Summary of $z\sim10$ LF and Luminosity Density Estimates\label{tab:summary}}
\tablewidth{0 pt}
\tablecolumns{5}
\tablehead{\colhead{$\log_{10}\phi_*(z=10)$} & \colhead{$M_*(z=10)$}&\colhead{$\alpha$} & \colhead{$\log_{10}\rho_L$\tablenotemark{$\dagger$} } &\colhead{$\log_{10}\rho_{*}$}  \\
$[$Mpc$^{-3}$mag$^{-1}]$ & &  &$[$erg$~$s$^{-1}$Hz$^{-1}$Mpc$^{-3}]$ & $[$\msol yr$^{-1}$Mpc$^{-3}]$
}
\startdata
$-2.9$ (fixed) & $-18.0\pm0.5$ & $-1.73$ (fixed)  & $24.2\pm0.5$ & $-3.7\pm0.5$\\
$-4.0\pm0.5$ & $-19.63$ (fixed)  & $-1.73$ (fixed)  & $24.4\pm0.5$ & $-3.5\pm0.5$\\
\multicolumn{3}{c}{single candidate} & $24.1^{+0.5}_{-0.7}$ & $-3.8^{+0.5}_{-0.7}$
\enddata

\tablenotetext{$\dagger$}{Integrated down to $0.06L_{z=3}^*$ ($M_{1400}=-18$ mag)}

\end{deluxetable*}

\subsection{Evolution in Luminosity Density From $z\sim10$ to $z\sim8$}

The quantity most easily comparable between observations and simulations is the observed luminosity density ($\rho_L$) above a given limiting magnitude, which is shown in Figure \ref{fig:LDevol}. The LD from the $z\sim10$ candidate alone amounts to $\log_{10}\rho_L=24.1^{+0.5}_{-0.7}$ erg$~$s$^{-1}$Hz$^{-1}$Mpc$^{-3}$. However, a more realistic estimate of the luminosity density can be obtained from the two possible extrapolations of the UV LF we derived in the previous section, which include the contribution from galaxies at $M_{1400}<-19$ mag that are currently undetected. Assuming the LF evolution only occurs in the characteristic luminosity, we find for the $z\sim10$ luminosity density $\log_{10}\rho_L=24.2\pm0.5$ erg$~$s$^{-1}$Hz$^{-1}$Mpc$^{-3}$, while assuming the evolution to be driven by a normalization of the Schechter function only, we find $\log_{10}\rho_L=24.4\pm0.5$ erg$~$s$^{-1}$Hz$^{-1}$Mpc$^{-3}$. These different estimates are also summarized in Table \ref{tab:summary}.

Given that the observed luminosity density at $z\sim8$ is $\log_{10}\rho_L(z=8)=25.6$ erg$~$s$^{-1}$Hz$^{-1}$Mpc$^{-3}$ \citep[][]{Bouwens10c}, the inferred increase in luminosity density in the 170 Myr from $z\sim10$ to $z\sim8$ amounts to more than an order of magnitude. This is a factor $\geq4$ higher than what would have been inferred from the empirical relation for the UV LF evolution, which predicts an increase by only a factor $\sim3$. 

Note however, that such a rapid increase in luminosity density is actually predicted by many theoretical models. In Figure \ref{fig:LDevol} we also show the  luminosity densities at $M_{1400}<-18$ mag as derived from the semi-analytical model (SAM) of \citet{Lacey11}, and from the theoretical model of \citet{Trenti10}. 

Although there is still some discrepancy on the exact shape of the $z>6$ UV LF between the \citet{Lacey11} SAM  and the observations, the integrated luminosity density and its evolution is remarkably well reproduced across the full redshift range $z\sim4-8$ \citep[see also discussion in][]{Raicevic11}. The semi-analytic model predicts a constant growth in $\rho_L$ with cosmic time at somewhat faster pace than observed, leading to some discrepancy between the observations and the model at $z\sim7$ and $z\sim8$, where the observational data shows higher luminosity densities. However, the model reproduces  our estimates of the $z\sim10$ LD remarkably well.

Similar conclusions are reached for the purely theoretical model of \citet{Trenti10}. In particular, this model predicts the UV LD to evolve at an accelerated rate at $z>8$, being only $\log_{10}\rho_L=24.4$ erg$~$s$^{-1}$Hz$^{-1}$Mpc$^{-3}$ at $z\sim10$, in excellent agreement with our observed estimates. Since the model is only based on the evolution of the underlying dark matter mass function, this indicates that an accelerated evolution in the galaxy population can be explained even without the need for a change in the physical mechanisms of galaxy formation.

For further theoretical model predictions, see also e.g. \citet{Mao07,Salvaterra11}, or \citet{Munoz11}.

It is also interesting to note that the star-formation rate densities (see Table \ref{tab:summary}) inferred from our data, are more than an order of magnitude too low to account for the stellar mass densities observed at $z\sim7$ in systems of similar brightness. With constant star-formation over $\sim300$ Myr from $z\sim10$ to $z\sim7$ the observed galaxy population would only produce a stellar mass density of $\log_{10}\rho_M=4.7-5.3~ M_\odot~$Mpc$^{-3}$, compared to $\log_{10}\rho_M(z=7)=6.6 ~M_\odot~$Mpc$^{-3}$ as estimated by, e.g., \citet[][]{Labbe10a,Labbe10b,Gonzalez11}.
If the inferred mass densities and SFRs are correct, this suggests that the majority of the stars found in $z\sim7$ galaxies down to $M_{UV}<-18$ mag have to be formed in systems below our detection limit at $z\sim10$ or are younger than 300 Myr.

\section{Summary and Conclusions}
\label{sec:discussion2}

In this paper, we have extended our search for $z\sim10$ galaxies to $\sim160$ arcmin$^2$ of public WFC3/IR data obtained around the GOODS South field (see Figure \ref{fig:fields}). These data sets have been acquired through the three surveys HUDF09, ERS, and CANDELS, and reach to varying depths, from $H_{160,AB}=26.9$ to $H_{160,AB}=29.4$. Based on strict optical non-detection requirements and a color cut of $J_{125}-H_{160}>1.2$, we search these fields for Lyman Break $J_{125}$-dropout galaxies, which are expected to lie at $z>9.5$. A total of 17 sources satisfy these criteria. However, 16 out of these sources show strong IRAC detections which rule out their being at such very high redshifts. Rather, these galaxies are found to have best-fit photometric redshifts in the range $z_{phot}=2-4$ (see section \ref{sec:contaminants} and appendix). They remain undetected in the optical due to their evolved stellar populations with non-negligible dust obscuration.
This shows how important Spitzer IRAC data is for removing contaminating lower redshift galaxies. 

Interestingly, these contaminants are essentially only detected with magnitudes in the range $H_{160,AB}=24-26$ mag. This is $\sim1$ mag brighter than the detection limits of our bright surveys, which suggests that such evolved and dusty galaxies follow a peaked luminosity function at these wavelengths. If confirmed by future wide-area WFC3/IR data sets, this would indicate that such extremely red systems are not as much of a problem for $z>9$ searches at fainter levels as has been expected to date.
However, the existence of such galaxy populations  will make it challenging to use large-area WFC3 surveys such as pure-parallel fields \citep[e.g.][]{Trenti11} for constraining the bright end of the $z\sim10$ UV LF without additional follow-up observations to validate the candidates.

Even with our expanded search area, the only $>5\sigma$ detected galaxy with a color limit of $H_{160}-[3.6]<1.6$ ($2\sigma$) and thus the only possible $z\sim10$ galaxy candidate is the same source that we reported  already in \citet{Bouwens11}. 
In appendix \ref{sec:possiblez10} we additionally note one other lower S/N candidate that is suggestive of
being at comparable redshift, but requires confirmation from further HST
WFC3/IR data.

Interestingly, we would have expected to detect six $z\sim10$ galaxies in our data, if the UV LF evolved to $z\sim10$ as expected from lower redshift trends (see \S \ref{sec:Nexp}). Thus, the galaxy population appears to evolve at an accelerated rate beyond $z>8$. We infer that the UV luminosity density increases by more than an order of magnitude in only 170 Myr from $z\sim10$ to $z\sim8$, and we are thus likely witnessing the first massive build-up of the galaxy population at these early epochs in the reionization era. The fact that theoretical models based on the evolution of dark matter halos do, in fact, predict such an accelerated increase in the LD, indicates that these rapid changes are mainly driven by an accelerated evolution of the underlying dark matter mass function, rather than due to a change in star-formation properties of these early galaxies \citep[see e.g.,][]{Trenti10,Lacey11}.

The accelerated evolution of the galaxy population also has interesting consequences for cosmic reionization by galaxies brighter than $M_{UV}=-18$ mag, the current detection limits. With such a sharp decrease in the luminosity density above $z\sim8$, it is impossible for such bright galaxies alone to create a reionization history in agreement with the high optical depth measurement of WMAP and the vast majority of the ionizing flux has to be created by fainter galaxies \citep[see also][]{Bouwens11b}.

\begin{figure}[tbp]
	\centering
	\includegraphics[width=\linewidth]{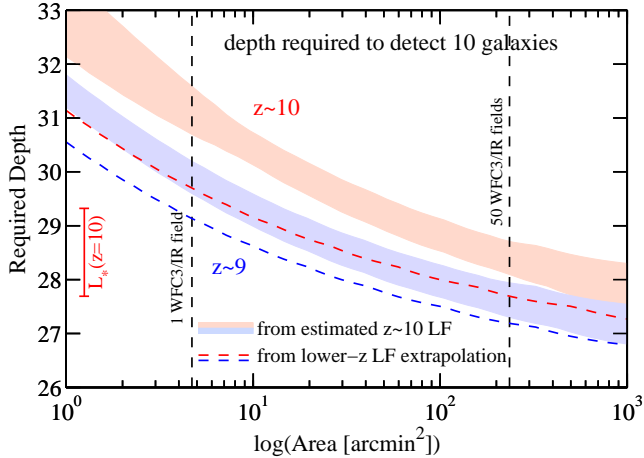}
  \caption{Depth required to detect ten $z\sim9$ galaxies (blue shaded region and dashed line) and ten $z\sim10$ galaxies (red shaded region and dashed line) as a function of survey area. These regions show the expected number of sources calculated using our LF extrapolations to $z\sim10$ from section \ref{sec:LFconstraints}, i.e. assuming only evolution in $M_*$ or only in $\phi_*$ to reproduce the detection of our one $z\sim10$ candidate. The dashed lines show the same, but assuming the standard lower-redshift extrapolation of the UV LF \citep[][]{Bouwens10c}. In particular, at $z\sim10$ the estimated depths differ significantly. The range of $L_*(z=10)$ values of these extrapolations is indicated with an errorbar in the lower left. As can be seen, surveys reaching significantly deeper than $\sim28$ mag will be required to detect a significant population of $z>8$ galaxies. The vertical dashed lines indicate the area of one and 50 WFC3/IR fields, respectively. A survey with only one pointing would need to reach to $\sim31$ mag in $H_{160,AB}$ to significantly constrain the $z\sim10$ galaxy population, which is out of reach with HST WFC3/IR. Therefore, multiple fields are favorable for searching for $z\sim10$ galaxies. Additionally, comparably deep optical data ($>30$ mag) and deep IRAC imaging would be required over such fields in order to robustly exclude low-redshift contaminants. The identification of $z\sim9$ galaxies would benefit from imaging in different filters (e.g., F140W) than adopted in current deep WFC3/IR fields.}
	\label{fig:Nrequired}
\end{figure}

Unfortunately, our knowledge of galaxies at $z>8$ still remains very uncertain. However,  WFC3/IR offers unique opportunities to make significant progress in expanding the number of galaxies at $z\sim9-10$ and addressing
some of the key issues related to early galaxy formation and its impact on
reionization, even before the advent of JWST. In Figure \ref{fig:Nrequired}, we show the depth required to detect ten $z\sim9$ and ten $z\sim10$ galaxies for a given survey area. Such future data sets will have to reach significantly fainter than $\sim28$ mag to accomplish this goal, even for large surveyed areas of 50 WFC3/IR fields. This is deeper than currently planned wide-area WFC3/IR data (including MCT programs). 
In fact, to
characterize the luminosity function at luminosities below L$_*$ and to set
better constraints for reionization, surveys to fainter than 29 AB mag are
really needed.
Note that with only one WFC3/IR field, a magnitude of $\sim31$  has to be reached to detect a significant $z\sim10$ population, which is fainter than is practical with HST. 

Based on the current LF constraints, we find that imaging a few fields provides the best chance to improve on current $z\sim10$ constraints.
Furthermore, in order to further constrain the possible accelerated evolution of the UV LF, the $z\sim9$ regime offers the best opportunity. A sizable population of $z\sim9$ galaxies is expected to be seen already down to $\sim29$ mag over multiple WFC3/IR fields, which can be achieved with the efficient F140W filter.
Thus, it is likely that already with WFC3/IR, we can soon push the frontier of statistical galaxy samples with WFC3/IR from $z\sim8$ another $\sim100-170$ Myr back out to $z\sim9-10$.

\acknowledgments{Support for this work was provided by NASA through Hubble Fellowship grant HF-51278.01 awarded by the Space Telescope Science Institute, which is operated by the Association of Universities for Research in Astronomy, Inc., for NASA, under contract NAS 5- 26555. This work has been supported by NASA grant NAG5-7697 and NASA grant HST-GO-11563.01. }

Facilities: \facility{HST(ACS/WFC3), Spitzer(IRAC)}.


\appendix

\section{The Intermediate Redshift Contaminants}
Here we briefly summarize the properties of the galaxies that  formally met the $J_{125}$-dropout criteria based on the HST data alone, but which were found to exhibit very red $H_{160}$ to IRAC colors. This makes it extremely unlikely that these galaxies are at $z\gtrsim9$. A summary of these sources is listed in Table \ref{tab:photContamin}. 

\begin{figure}[tbp]
	\centering
	\includegraphics[scale=0.65]{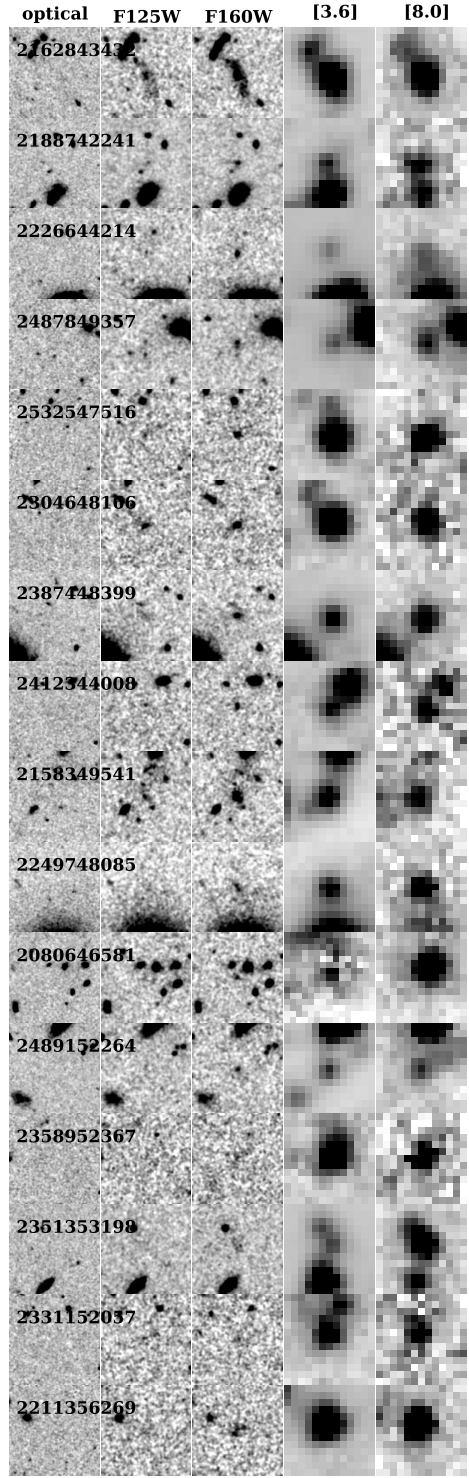}
  \caption{Images of all optical undetected sources with $J_{125}-H_{160}>1.2$ mag, but with strong IRAC detections. This rules out the possibility that these sources are at $z>9$. The images are 7.5 arcsec on a side, and show from left to right (1) a stack of $B_{435}$, $V_{606}$, $i_{775}$, $z_{850}$, (2) $J_{125}$, (3) $H_{160}$, (4) Spitzer IRAC [3.6], and (5) IRAC [8.0]. All these sources are clearly detected in all IRAC bands, including [8.0]. Thus, their $H_{160}-[8.0]$ colors are greater than $\sim2$, which is very different from the relatively flat color expected for a real $z\sim10$ source.  Note that IRAC flux of the source jD-2080646581 is heavily contaminated by neighboring sources. In the [3.6] band, we therefore show the cleaned image, after subtraction of the contaminating flux, which reveals its clear detection. }
	\label{fig:stampsContamin}
\end{figure}

Note that all these galaxies are obviously clearly detected in the IRAC images, with the exception of one source (jD-2080646581), which is close to a quintet of IRAC bright galaxies. However, after subtraction of the contaminating flux from its neighbors based on a careful convolution of the $H_{160}$ image to the IRAC PSFs \citep[see, e.g.,][]{Labbe10a,Labbe10b,Gonzalez11} it also shows a clear detection (see Fig. \ref{fig:stampsContamin}). The best estimate for its color is  $H_{160}-[3.6]=1.6\pm0.5$, which is too red for a likely $z>9$ source.

Interestingly, six of these 16 sources are present in catalogs of potential passive high redshift galaxies of \citet{Rodighiero07} and \citet{Wiklind08}.
One of these is also detected in X-ray emission and classified as a so-called EXO source hosting a potentially highly obscured AGN, particularly if it is an evolved galaxy at intermediate redshift, as pointed out by \citet{Koekemoer04}.

We have derived photometric redshifts and mass estimates for these galaxies by complementing our HST photometry with the IRAC fluxes from the GOODS MUSIC catalog \citep{Santini09} or from the SIMPLE images. Based on SED fits with \citet{Bruzual03} models, these galaxies are confirmed to be evolved, reddened systems with stellar masses around $10^{11}$ \msol\ at redshifts $z\sim2-3.5$. Note that with the exception of four sources these are all detected individually in MIPS 24$\micron$ data.

\begin{figure}[bp]
	\centering
	\includegraphics[scale=0.65]{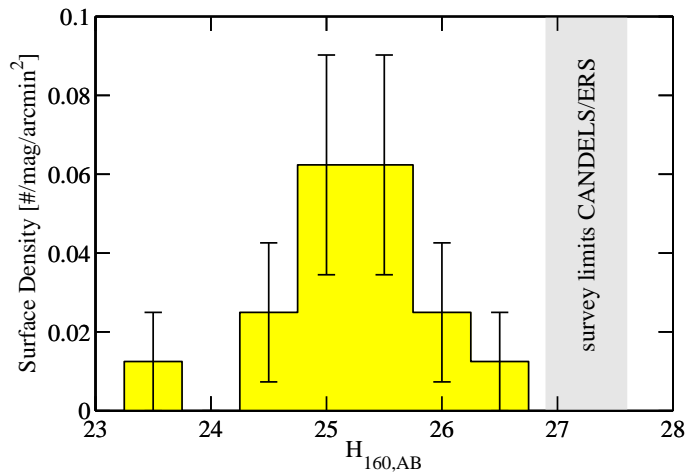}
  \caption{The surface density of contaminating sources at redshifts $z\sim2-4$. The surface densities are extremely low. Only $\sim$0.4 sources are expected per WFC3/IR pointing. The rapid disappearance of such sources beyond $H_{160,AB}\gtrsim26$, almost one mag brighter than the detection limits of our shallow fields from ERS and CANDELS, suggests that such sources may have a peaked LF at the wavelengths of
interest in these surveys and so could be less of a problem for contamination for fainter $z\sim10$ galaxy samples.}
	\label{fig:ContaminMz}
\end{figure}

In Figure \ref{fig:ContaminMz} we additionally show the surface density of these evolved intermediate redshift sources. Even at their peak surface density, they are only found at 0.06 sources per magnitude per arcmin$^2$. Therefore, only $\sim0.4$ sources are expected per WFC3/IR pointing, providing an explanation for why none of these bright sources are found in the three deep HUDF09 fields. The dearth of such galaxies faintward of $H_{160,AB}=26$, at $\sim1$ mag brighter than the completeness limit of our bright surveys, suggests that such extremely red sources become even less frequent (i.e., that their LF may have peaked at these magnitudes) and so they could be less problematic for contamination of fainter $z\gtrsim9$ dropout searches. This will have to be confirmed, however, with future wide-area WFC3/IR data, searching explicitly for this type of galaxies.

Note that our finding of a disappearance of passive intermediate redshift galaxies to fainter magnitudes is in agreement with \citet{Stutz08}, who found an absence of passive, red galaxies at $z\sim1.5-3$ at masses below $\sim10^{10}$ \msol.

\begin{deluxetable}{lcccccccccc}
\tablecaption{Lower Redshift Contaminants that satisfy the J-Dropout criteria, but show strong IRAC detections.\label{tab:photContamin}}
\tablewidth{0 pt}
\tablecolumns{9}
\tabletypesize{\scriptsize}
\tablehead{\colhead{ID} & $\alpha$ & $\delta$ &\colhead{$H_{160}$}  &\colhead{$J_{125}-H_{160}$} &  \colhead{S/N$_{H160}$}   & \colhead{$r_{1/2}[\arcsec]$} & \colhead{MUSIC-ID\tablenotemark{$\dagger$}} & \colhead{references} }

\startdata

 \multicolumn{9}{c}{\it ERS}\\[0.15cm]
jD-2162843432 &  03:32:16.28 &  -27:43:43.2 &  $23.6\pm0.1$ & $1.32\pm0.11$ &  14.6  & 0.65 & 70081 & --    \\
jD-2188742241 &  03:32:18.87 &  -27:42:24.1 &  $25.5\pm0.1$ & $1.36\pm0.23$ &  12.2  & 0.29 & 70040 & --    \\  
jD-2226644214 &  03:32:22.66 &  -27:44:21.4 &  $25.5\pm0.1$ & $1.80\pm0.38$ &  10.9  & 0.26  & 70104 & --  \\[0.15cm]

\multicolumn{9}{c}{\it CANDELS-Deep}\\[0.15cm]
jD-2487849357  &  03:32:48.78  &  -27:49:35.7  &  $25.2\pm0.1$ & $1.36\pm0.25$  &  11.4  & 0.33 & 70314 & --    \\
jD-2532547516  &  03:32:53.25  &  -27:47:51.6  &  $25.3\pm0.2$ & $1.55\pm0.32$  &  11.2  & 0.26&70236 & 4  \\
jD-2304648166  &  03:32:30.46  &  -27:48:16.6  &  $24.8\pm0.1$ & $1.23\pm0.11$  &  29.4  & 0.23 & 70258 & 1 \\ 
jD-2387448399  &  03:32:38.74  &  -27:48:39.9  &  $24.9\pm0.1$ & $1.41\pm0.22$  &  14.9  & 0.33    & 70273 & 1,2,4 \\ 
jD-2412344008  &  03:32:41.23  &  -27:44:00.8  &  $25.8\pm0.2$ & $1.43\pm0.30$  &  11.0  & 0.22    & 70092 & --   \\
jD-2158349541  &  03:32:15.83  &  -27:49:54.1  &  $24.4\pm0.1$ & $1.60\pm0.15$  &  19.1  & 0.38    & 70316 & 1 \\
jD-2249748085  &  03:32:24.97  &  -27:48:08.5  &  $25.7\pm0.2$ & $1.61\pm0.43$  &  7.5  & 0.33    & 70252 & --    \\
jD-2080646581  &  03:32:08.06  &  -27:46:58.1  &  $26.5\pm0.3$ & $2.01\pm0.67$  &  6.1  & 0.16  & -- & --   \\[0.15cm]

\multicolumn{9}{c}{\it CANDELS-Wide}\\[0.15cm]
jD-2489152264 &  03:32:48.91 &  -27:52:26.4 &  $25.2\pm0.1$ & $1.49\pm0.27$ &  12.3  & 0.26 & 70442 & --    \\
jD-2358952367 &  03:32:35.89 &  -27:52:36.7 &  $25.1\pm0.2$ & $2.27\pm0.94$ &  7.2  & 0.35  & 70455 & 1  \\
jD-2351353198 &  03:32:35.13 &  -27:53:19.8 &  $25.4\pm0.2$ & $1.61\pm0.43$ &  5.8  & 0.35  & 70484 & --   \\
jD-2331152057 &  03:32:33.11 &  -27:52:05.7 &  $26.2\pm0.3$ & $1.24\pm0.54$ &  5.8  & 0.20  & 70429 & 1,3  \\  
jD-2211356269 &  03:32:21.13 &  -27:56:26.9 &  $24.4\pm0.1$ & $1.47\pm0.22$ &  7.9  & 0.44   & -- & -- 

\enddata

\tablenotetext{$\dagger$}{From \citet{Santini09}.}

\tablerefs{(1) \citet{Rodighiero07}; (2) \citet{Mobasher05}; (3) \citet{Koekemoer04}; (4) \citet{Wiklind08}}
\end{deluxetable}

\section{Potential Future Extension of the HUDF $z\sim10$ Candidate Sample}
\label{sec:possiblez10}

As pointed out in section \ref{sec:discussion2}, due to the apparent paucity of such sources, it will be difficult to extend the $z\sim10$ candidate lists to large numbers of reliable sources in the near future with HST. In cycle 18, an additional 128 orbit imaging survey was granted (PI: Ellis) to further study the $z\gtrsim7$ galaxy population over the HUDF data. However, given that only one single field will be targeted and that only a relatively small amount of $H_{160}$ data is currently planned to be taken (22 new orbits, relative to the $\sim$58 already available from the HUDF09 and CANDELS programs), we can forecast what this program will bring in terms of $z\sim10$ galaxy science. Essentially, every 5$\sigma$ source that will be obtained as a result of this new program should already be present as a 4.2$\sigma$ source in the current $H_{160}$ band data. We therefore systematically searched for lower significance sources (down to 4$\sigma$ detections) that satisfy our $z\sim10$ dropout criteria. Essentially, the only additional candidate that we found is already detected at 4.8$\sigma$ in the current data, and is located at RA$=$03:32:43.01, DEC$=$-27:46:53.3 \citep[see also][]{Yan10}. 
Note, however, that the $i_{775}$-band data shows very faint positive flux ($\sim1.5\sigma$) exactly at the location of this source, which increases the chance that it is at lower redshift. This demonstrates the value of even deeper optical data over these fields for more robust high redshift galaxy selections.
Given the scarcity of low significance candidates in the HUDF, it is clear, however, that it will be very important to image at least one additional field to similar depth in order to further constrain the accelerated evolution in the cosmic star-formation rate density (see Fig. \ref{fig:Nrequired}).

\end{document}